\documentclass[twocolumn]{revtex4}

\newcommand{\smsb}[1]{\mbox{\tiny #1}}
\newcommand{\fig}[2]{\includegraphics[width=#1]{./#2.eps}}
\usepackage{graphicx}

\begin{document}

\title{Exchange Bias with Interacting Random Anti-ferromagnetic Grains}

\author{Hwee Kuan Lee and Yutaka Okabe}

\affiliation{Department of Physics, Tokyo Metropolitan University, Hachioji,
  Tokyo 192-0397, Japan}

\begin{abstract}
A model consisting of random interacting anti-ferromagnetic (AF) grains 
coupled to a ferromagnetic (FM) layer is developed to study the exchange bias
phenomenon. This simple model is able to describe several exchange bias 
behavior observed in real materials. Shifts in hysteresis loops are observed 
as a function of cooling field and average grain size. We establish a 
direct relationship between cooling field dependence of exchange bias, 
coercivity and magnetization state on the AF-FM interface. We also verify that
the exchange bias field is inversely proportional to the grain size, and this
behavior is independent of the inter-grain interactions, AF/FM coupling and 
cooling field.
\end{abstract}

\maketitle
%%%%%%%%%%%%%%%%%%%%%%%%%%%%%%%%%%%%%%%%%%%%%%%%%%%%%%%%%%%%%%%%%%%%%%%

%
%\section{Background}
%

When a ferromagnetic (FM) material is coupled to an anti-ferromagnetic (AF) 
material, under suitable conditions, a unidirectional anisotropy is observed.
This results in a shift in the hysteresis loop called 
{\it exchange bias}~\cite{meiklejohn,nogues}. Because of its application to 
spin valves, exchange bias has been studied extensively, but the roles of many
parameters, such as magnetic domains and cooling field, have not been fully 
understood.
There have been attempts to understand exchange bias in the microscopic 
level~\cite{nowak,almeida}. 
In particular, recent publications on the domain state model~\cite{beckmann} 
argued that exchange bias is due to different domain 
orientations in the AF bulk. In this paper, we make a theoretical analysis 
based on a microscopic model to have a better understanding on the roles of 
domain orientations. In contrast with previous works~\cite{beckmann,scholten}, 
where domains were introduced by dilution, we model domains explicitly with 
grains on the AF materials. We derive a set of mean-field equations that 
consider a free boundary on the surface of each AF grain and an effective 
field on the surface due to interactions with other grains. We study the 
cooling field and grain size dependences of exchange bias.

%
%\section{Model}
%

\begin{figure}
\fig{6cm}{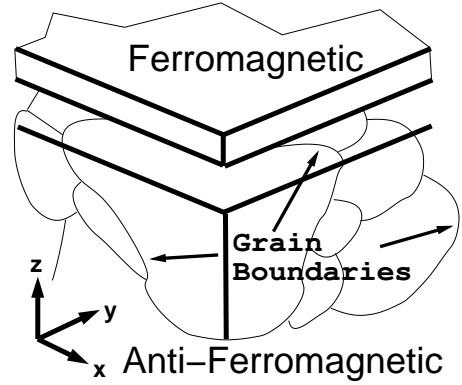}
\caption{A schematic of FM layer with AF grains}
\label{fig:schematic}
\end{figure}

Our model for studying exchange bias consists of {\it one} FM layer on a square
lattice coupled to multiple AF layers on a cubic lattice. We used {\it eight} 
AF layers in all our calculations. 
Fig.~\ref{fig:schematic} shows a schematic of the AF grains 
and defines the coordinate system which we shall use hereafter. 
Periodic boundary conditions are used for in-plane directions, 
and free boundary conditions are considered in the $z$ direction. 
The Hamiltonian for the FM layer is given by
\begin{equation}
{\cal H}_{\smsb{FM}} = 
  -J_{\smsb{FM}} \sum_{\langle ij\rangle} \vec{s}_i . \vec{s}_j - 
  \sum_i \left[  d ( s_i^x )^2 + h s_i^x + 
  J_{\smsb{I}}  s_i^x \sigma_i \right]
\end{equation}
where $\vec{s}_i$ are Heisenberg spins 
and $J_{\smsb{FM}}$ is the exchange coupling between FM spins.  
The first sum is performed over nearest neighbors $\langle ij\rangle$ 
within the FM layer. 
The $i$ sum is taken over all lattice sites. 
The uniform external field $h$ is chosen to be parallel 
to the easy axis x, and $d = 0.1 J_{\smsb{FM}}$ is the anisotropy constant. 
$J_{\smsb{I}}$ is the exchange coupling between FM and AF layers, 
and several different values of $J_{\smsb{I}}$ are used in our calculations. 
For spins of AF layers we take Ising spins $\sigma_i = \pm 1$.
% $J_{\smsb{I}}$ is the exchange coupling between FM and AF
% where several different values are used in our calculations. 
% $\sigma_i = \pm 1$ are the spins of the top AF layer.
%
The Hamiltonian for one AF grain is 
\begin{equation}
{\cal H}_{\smsb{grain}} = 
  - J_{\smsb{AF}} \sum_{\langle i j \rangle} \sigma_i \sigma_j 
  -  h \sum_i \sigma_i
  - \sum_{n}^{\smsb{surface}} h^{\smsb{sur}}_n \sigma_n
\end{equation}
where the coupling $J_{\smsb{AF}}$ is chosen as  $- 0.5 J_{\smsb{FM}}$. 
The first and second sums are taken over the nearest neighbors 
and all lattice sites of each grain, respectively. 
$h_n^{\smsb{sur}}$ is an additional effective
field on the surface, and the $n$ sum is performed over the surface 
of the grain.  The effective field $h_n^{\smsb{sur}}$ consists of two terms. 
The first term is the contribution from the nearest neighbor spins belonging 
to different grains, $J_{\smsb{g}} \sum_j \sigma_j$, where we take
$J_{\smsb{g}} = -0.15 J_{\smsb{FM}}$ ($|J_{\smsb{g}}| < |J_{\smsb{AF}}|$). 
The second term is $J_{\smsb{I}} s_k^x$ if the site $n$ belongs to the 
interface layer, and $k$ is the nearest neighbor site of $n$ in the FM layer.

%--------------------------------------
%
%--------------------------------------
%
%--------------------------------------
\begin{figure}
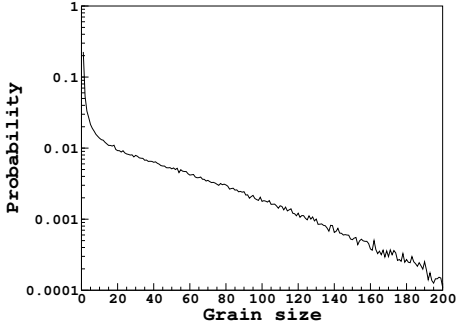

\fig{6cm}{grainsize_dist}
\caption{Plot of grain size 
distribution with mean grain size = 32.}
\label{fig:grains}
\end{figure}
%--------------------------------------
%

%%%%%%%%%%%%%%%%%%%%%%%%%%%%%%%%%
% begin revise 
%%%%%%%%%%%%%%%%%%%%%%%%%%%%%%%%%
To generate the grain configuration, we have take the following procedure.
We start from an ``unoccupied" 
AF lattice and insert $n_{\smsb{g}}$ seeds, where
each seed is the starting point of a distinct grain. 
From these seeds, we grow grains by the following process.
Randomly choose a site.  If an empty site is chosen, choose another site.
If the chosen site belongs to a grain, let all empty 
neighbors of this site belong to the same grain.  Repeat this step 
until all sites are occupied. 
%%%%%%%%%%%%%%%%%%%%%%%%%%%%%%%%%
% end revise 
%%%%%%%%%%%%%%%%%%%%%%%%%%%%%%%%%
%
%
%
Fig.~\ref{fig:grains} shows the grain size distribution with 512 grains on 
$32 \times 64\times 8$ lattice averaged over $512$ samples. Grain 
configurations with $512$ grains and average grain size of 32 lattice sites are
used unless otherwise stated.
%
%

%
%\section{Mean-field Theory}
%

We derive a set of coupled mean-field equations for the magnetization of each AF 
grain. These equations account for the free surfaces and the surface fields due 
to interactions with other grains. Details of derivations will be shown 
elsewhere~\cite{hkl}. For the $j$th grain, it follows that
%
%%%%%%%%%%%%%%%%%%%%%%%%%%%%%%%%%%%%%%%%%%%%%%%%%%%%%%%%%%%%%%%%%%%%%%%
%
% equation for AF magnetization
%
%%%%%%%%%%%%%%%%%%%%%%%%%%%%%%%%%%%%%%%%%%%%%%%%%%%%%%%%%%%%%%%%%%%%%%%
\begin{eqnarray}
\lefteqn{%
m^\chi_j 
  \left[ 
  \frac{q}{2} ( \rho_{Bj}^\psi + \rho_{Bj}^\chi ) + 
  \frac{q_{\smsb{S}}}{2} ( \rho_{Sj}^\psi + \rho_{Sj}^\chi ) 
  \right] = } \nonumber \\
& & 
\rho_{Bj}^\chi q \tanh[\beta ( J_{\smsb{AF}} q m^\psi_j + h) ] + \nonumber \\
& &
\rho_{S \smsb{free} j}^\chi q_{\smsb{S}} 
\tanh[ \beta ( J_{\smsb{AF}} q_{\smsb{S}}
m^\psi_j + h )] + \nonumber \\ 
& &
\rho_{S \smsb{FM} j}^\chi q_{\smsb{S}} 
\tanh[ \beta ( J_{\smsb{AF}} q_{\smsb{S}}
m^\psi_j + h + J_{\smsb{I}} M^x )] + \nonumber \\ 
& & 
\sum_{k=1}^{n_{\smsb{g}}} \rho_{Sjk}^\chi  q_{\smsb{S}}
\tanh[ \beta ( J_{\smsb{AF}} q_{\smsb{S}} m^\psi_j + h + 
J_{\smsb{g}} q_{\smsb{g}} m^\psi_k) ]  
\label{eq:maf}
\end{eqnarray}
%%%%%%%%%%%%%%%%%%%%%%%%%%%%%%%%%%%%%%%%%%%%%%%%%%%%%%%%%%%%%%%%%%%%%%%
%
%
%
where the variables $\psi$ and $\chi$ are used to
represent sublattices, that is, $m^\psi_j$ and $m^\chi_j$ 
are the magnetization of one sublattice of the $j$th grain 
and that of the other sublattice, respectively.  
For each grain, there are a pair of symmetric 
equations for both sublattices.  
As a total, Eq.~(\ref{eq:maf}) consists of a set of $2n_{\smsb{g}}$ equations. 
The variables in Eq.~(\ref{eq:maf}) are defined as
follows:
%
%
%%%%%%%%%%%%%%%%%%%%%%%%%%%%%%%%%%%%%%%%%%%%%%%%%%%%%%%%%%%%%%%%%%%%%%%
%
% definitions for variables
%
%%%%%%%%%%%%%%%%%%%%%%%%%%%%%%%%%%%%%%%%%%%%%%%%%%%%%%%%%%%%%%%%%%%%%%%
$q = 6$                     is the coordination number in the bulk,
$q_{\smsb{S}} $             is the average coordination number on the surface,
$q_{\smsb{g}}$              is the average coordination number between sites 
                            belonging to different grains, 
and $\beta$                 is the inverse temperature.  
By defining the densities of bulk/surface sites
as the number of bulk/surface sites divided by the
total number of sites in the simulation box, we express the mean-field
equation in terms of densities.
$\rho_{Bj}^{\psi/\chi}$ and $\rho_{Sj}^{\psi/\chi}$ are 
    the densities of bulk sites and surface sites for sublattice
                            $\psi / \chi$, respectively.
$\rho_{S\smsb{free}j}^\chi$ is the density of free surface, 
$\rho_{S\smsb{FM}j}^\chi$   is the density of sites adjacent to the FM layer,
and
$\rho_{Sjk}^{\psi}$         is the density of surface sites adjacent 
                            to the neighboring $k$th grain.
%%%%%%%%%%%%%%%%%%%%%%%%%%%%%%%%%%%%%%%%%%%%%%%%%%%%%%%%%%%%%%%%%%%%%%%
%
%
The first term in the right hand side of Eq.(\ref{eq:maf}) gives the 
contribution to $m^\chi_j$ due to magnetization ordering in the bulk.
The second term gives the contribution due to the free surfaces at $z=0$.
The third term is due to interactions with the top FM layer and the last
term is due to inter-grain interactions.
For the FM layer the free energy is calculated as
%
%%%%%%%%%%%%%%%%%%%%%%%%%%%%%%%%%%%%%%%%%%%%%%%%%%%%%%%%%%%%%%%%%%%%%%%
%
%  free energy for FM layer
%
%%%%%%%%%%%%%%%%%%%%%%%%%%%%%%%%%%%%%%%%%%%%%%%%%%%%%%%%%%%%%%%%%%%%%%%
\begin{eqnarray} 
\label{eq:fmf}
f & = & 
\frac{J_{\smsb{FM}} q_{\smsb{FM}} M^2 }{2}  \\ \nonumber
&& -\frac{1}{\beta} \sum_{k}^{n_{\smsb{g}}} 
  \left( 
  \rho_{S \smsb{FM} k}^\psi \log ( I[\vec{M},\vec{h}_k^\psi] ) +
  \rho_{S \smsb{FM} k}^\chi \log ( I[\vec{M},\vec{h}_k^\chi] ) 
  \right) 
\end{eqnarray}
%
%\begin{eqnarray}
%f & = & 
%\frac{J_{\smsb{FM}} q_{\smsb{FM}} M^2 }{2} - \nonumber \\
%%
%& &
%%
%\frac{1}{\beta} \sum_{k}^{n_{\smsb{g}}} 
%  \left( 
%  \rho_{S \smsb{FM} k}^\psi \log ( I[\vec{M},\vec{h}_k^\psi] ) +
%  \rho_{S \smsb{FM} k}^\chi \log ( I[\vec{M},\vec{h}_k^\chi] ) 
%  \right) \nonumber  \\
%\label{eq:fmf}
%\end{eqnarray}
%
%%%%%%%%%%%%%%%%%%%%%%%%%%%%%%%%%%%%%%%%%%%%%%%%%%%%%%%%%%%%%%%%%%%%%%%
%
%
where $\vec{M}$ is the FM  magnetization, $q_{\smsb{FM}}=4$ is the 
coordination number of FM layer, $\vec{h}_k^{\psi/\chi}$ is an effective field
defined by
$\vec{h}_k^{\psi/\chi} = ( h + J_{\smsb{I}} m_k^{\psi/\chi} ) \hat{e}_x$.
$I[\vec{M},\vec{h}_k^{\psi/\chi}]$ is an integral to be evaluated numerically,
%
%
%%%%%%%%%%%%%%%%%%%%%%%%%%%%%%%%%%%%%%%%%%%%%%%%%%%%%%%%%%%%%%%%%%%%%%%
%
% integral for single particle partition function
%
%%%%%%%%%%%%%%%%%%%%%%%%%%%%%%%%%%%%%%%%%%%%%%%%%%%%%%%%%%%%%%%%%%%%%%%
\begin{equation}
I[\vec{M},\vec{h}] = 
  \int_{4 \pi} d\vec{s} \exp \left[ \beta ( J_{\smsb{FM}}
  q_{\smsb{FM}} \vec{M} + \vec{h} ) \cdot \vec{s} + 
  \beta d (s^x)^2 \right]
\end{equation}
%%%%%%%%%%%%%%%%%%%%%%%%%%%%%%%%%%%%%%%%%%%%%%%%%%%%%%%%%%%%%%%%%%%%%%%
%
%
The $x$ component of $\vec{M}$, $M^x$, appears in Eq.~(\ref{eq:maf}).
Computation of mean-field magnetizations consists of solving Eq.~(\ref{eq:maf}) 
iteratively and simultaneously minimizing Eq.~(\ref{eq:fmf}).

Let us explain the field-cool process.  From now on, 
we use dimensionless units with $J_{\smsb{FM}}=1$, and 
the temperature $T$ is given in units of $J_{\smsb{FM}}/k_{\mbox{\tiny B}}$, 
where $k_{\mbox{\tiny B}}$ is the Boltzmann constant. 
The system is first cooled under the field $h_{\smsb{cool}}$
from $T=5$ to $T=0.2$ in steps of $T_{i+1} = 0.95 T_i$. 
There are two possible domain orientations for the AF grains, with sublattice 
$\psi$ pointing in the $+1$ direction and $\chi$ pointing in the $-1$ 
direction, or vice versa. During the field-cool process, we attempt to 
obtain the
global free energy minimum by considering two different domain orientations 
for each AF grain and pick up the one with lower free energy. 
At the end of the field-cool process, the temperature is kept constant and the
field $h$ is reduced in steps of 0.01 and with a higher resolution near the region
of magnetization reversal (0.002). At each step, the magnetization is 
calculated using the previous magnetization state as the starting values for
Eq.~(\ref{eq:maf}) and Eq.~(\ref{eq:fmf}). In this way, meta-stable states can
be traced out to obtain the hysteresis loop.  We simulate the field-cool 
process and the hysteresis loop based on the mean-field equations. 

%
%\section{Results}
%
%
%--------------------------------------
\begin{figure}
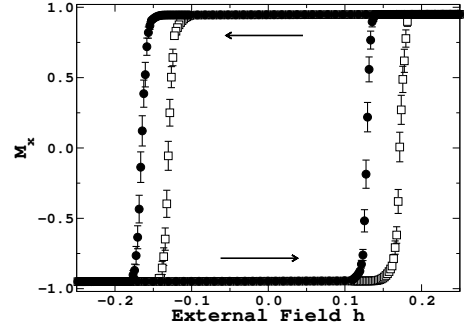

\fig{6cm}{hyst}
\caption{Hysteresis loop for $J_{\smsb{I}}=-0.5$ (white squares) and 
$J_{\smsb{I}}=0.5$ (black circles), with representative error bars obtained 
from 64 independent simulations.}
\label{fig:hyst}
\end{figure}
%--------------------------------------
%
%--------------------------------------
\begin{figure}
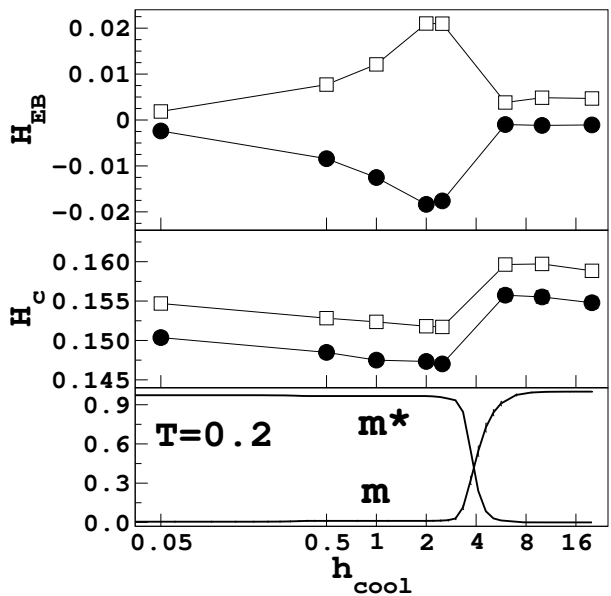

\fig{8cm}{cooling_fieldHebHc}
\caption{The dependence of $H_{\smsb{EB}}$ and $H_{\smsb{C}}$ on cooling field 
for $J_{\smsb{I}}=-0.5$ (white squares) and $J_{\smsb{I}}=0.5$ (black circles). 
The bottom plot shows the magnetization $m$ and staggered magnetization 
$m^*$ at the AF/FM interface at $T=0.2$.}
\label{fig:cf}
\end{figure}
%--------------------------------------
%

Fig.~\ref{fig:hyst} shows the hysteresis loops at $T=0.2$. Hysteresis loops 
shifted to the right for $J_{\smsb{I}} =-0.5$ (white squares) and left for 
$J_{\smsb{I}} = 0.5$ (black circles). Field-cooling is performed at 
$h_{\smsb{cool}} = 2$. 
%
%%%%%%%%%%%%%%%%%%%%%%%%%%%%%%%%%%%%%%%%%%%%%%%%%%%%%%%%%%%%%%%%%%%%%%%
%
% not so important part, need to shorten this part
%
%%%%%%%%%%%%%%%%%%%%%%%%%%%%%%%%%%%%%%%%%%%%%%%%%%%%%%%%%%%%%%%%%%%%%%%
%
%{\it 
%With a positive cooling field, excess magnetization in the AF bulk point in 
%the positive direction. Comparatively to the FM layer, it is rather hard to
%reverse the Ising spins in the AF bulk and most spins will be locked in their
%orientations during the field sweep of the hysteresis loop.
%When the AF-FM coupling is positive, these excess 
%magnetization holds the FM layer in positive direction resulting in a 
%hysteresis loop shift to the left. When the AF-FM coupling is negative the 
%hysteresis loop shifts to the right for the same reason. }
%%%%%%%%%%%%%%%%%%%%%%%%%%%%%%%%%%%%%%%%%%%%%%%%%%%%%%%%%%%%%%%%%%%%%%%
% replace by this short sentence
%
The origin of the shift comes from excess magnetization in the AF bulk
created during field-cooling.
%
%%%%%%%%%%%%%%%%%%%%%%%%%%%%%%%%%%%%%%%%%%%%%%%%%%%%%%%%%%%%%%%%%%%%%%%
This result is 
consistent with the intuitive picture discussed by Nogu\'{e}s and 
Schuller~\cite{nogues}. The shift in hysteresis
loop should depend on the magnetization state immediately before the 
hysteresis loop is traced out. We shall modify the magnetization state by 
using different cooling fields.

%
%--------------------------------------
\begin{figure}
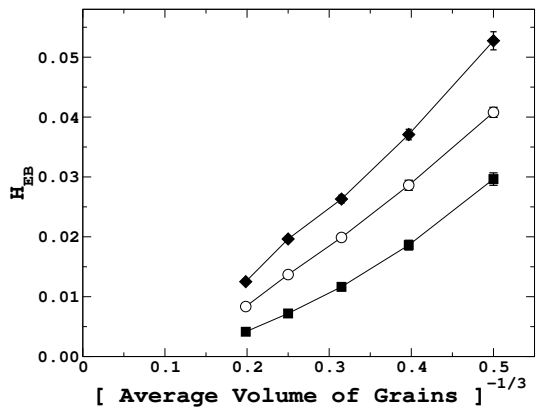

\fig{7cm}{grain_dep}
\caption{%
Grain size dependence of exchange bias for 
$J_I=-0.5, J_g=-0.15, h_{\smsb{cool}}=1.0$ (filled squares),
$J_I=-0.5, J_g=-0.15, h_{\smsb{cool}}=2.0$ (empty circles),
$J_I=-0.5, J_g=0, h_{\smsb{cool}}=1.0$ (filled diamonds).
Error bars are smaller than the size of the symbols.}
\label{fig:gd}
\end{figure}
%--------------------------------------
%

We performed a systematic study on the dependence of the exchange bias field 
$H_{\smsb{EB}}$ and coercivity $H_{\smsb{C}}$ in the field-cool process. 
$H_{\smsb{EB}}$ is defined as $(H_+ + H_-)/2$ where $H_+$ and $H_-$ are the 
fields at which the magnetization is zero in the hysteresis loop. Similarly, 
we define $H_{\smsb{C}} = (H_+ - H_-)/2$. Fig.~\ref{fig:cf} shows that 
$H_{\smsb{EB}}$ increases from zero to a maximum at about 
$h_{\smsb{cool}}=2.5$ and 
decreases at $h_{\smsb{cool}}>6$. The hysteresis loop remains unshifted with 
$h_{\smsb{cool}}=0$ when cooled from a demagnetized state; this observation is
consistent with experiments on exchange 
bias~\cite{meiklejohn,takahashi,steams}. Increase and decrease of 
$H_{\smsb{EB}}$ and $H_{\smsb{C}}$ with cooling field is also reported in real
materials such as Ni/NiFe$_2$O$_4$~\cite{negulescu} and 
permalloy/CoO~\cite{timothy,ambrose}.
As shown in the bottom plot of Fig.~\ref{fig:cf}, at $h<2.5$, the spins of 
AF-FM interface are locked in a magnetization state 
with staggered magnetization 
$m^* \approx 1$, and at $h>6$, the staggered magnetization vanishes. Our
result shows that cooling field of $h_{\smsb{cool}}=2.5$, for the sets of
parameters used in the present study, produces maximum excess 
magnetization in the AF-FM interface while maintaining non-zero staggered 
magnetization, causing a maximum exchange bias. At higher cooling field, spins
in the AF material are aligned during field-cooling; when the field is 
decreased to sweep the hysteresis loop, the AF grains get locked in AF states
with random domain orientations that generate little excess magnetization.
Hence there is a close relationship between the magnetization state of the
AF/FM interface and $H_{\smsb{EB}}$.
We also found, as in~\cite{timothy}, that there is a decrease of coercivity 
corresponding to an increase of exchange bias.
The dependence of exchange bias on cooling field is of particular importance in 
fabrication of exchange bias devices where tuning of exchange bias is 
desirable.

Fig.~\ref{fig:gd} shows grain size dependence of exchange bias. 
Average grain sizes of 8, 16, 32, 64 and 128 are used. Simulations 
with interacting/non-interacting grains and 
different cooling fields were performed. The exchange bias field is inversely
proportional to the grain diameter $D$, $H_{\smsb{EB}} \sim 1 / D$. For 
non-interacting grains (filled diamonds) and strong cooling field 
(empty circles), plots of exchange bias versus inverse grain diameter $D^{-1}$
fall on a straight line very well. For smaller cooling fields (filled 
squares), exchange bias goes to zero asymptotically at large grain sizes.
Simulations with $J_I=-0.75$ were also performed and the exchange bias field
changes only slightly compared to simulations with $J_I=-0.5$.
This inverse relationship of exchange bias on grains size is also reported in
exchange bias systems of permalloy/CoO bilayers~\cite{takano} and 
Cr$_{70}$Al$_{30}$/Fe$_{19}$Ni$_{81}$ bilayers~\cite{uyama}.
%

%%%%%%%%%%%%%%%%%%%%%%%%%%%%%%%%%
% begin revise 
%%%%%%%%%%%%%%%%%%%%%%%%%%%%%%%%%
Other models has been developed to describe the mechanism of unidirectional
anisotropy. Stiles and McMichael~\cite{stiles} used an ordered granular model
to explain exchange bias through partial domain wall formation. 
Scholten {\it et al}.~\cite{scholten} used mean-field equations on local
magnetization to explain the domain state model. The model presented here
is distinct from all previous models. The mean-field
equations for an explicit grain distribution has been given for the first time
in this paper, and the domain state model is realized explicitly by actual grain 
distribution.
%%%%%%%%%%%%%%%%%%%%%%%%%%%%%%%%%
% end revise 
%%%%%%%%%%%%%%%%%%%%%%%%%%%%%%%%%

%
%\section{Discussions}
%

To summarize, we developed a simple model that captures many features 
of real exchange bias
systems. We found a direct relationship between the cooling field dependence of
the exchange bias, coercivity and magnetization states on the AF-FM interface.
We also verified that the exchange bias field is inversely proportional to the
AF grain sizes and this relationship is independent of the inter-grain 
interactions, AF/FM interactions and cooling fields. Lastly, we would like to
mention that the simulations based on mean-field equations used in this paper
are general, and may be used to study other exchange bias parameters, 
such as surface roughness, perpendicular coupling and film thickness.

This work is supported by a Grant-in-Aid for Scientific Research from the 
Japan Society for the Promotion of Science. The computation of this work has 
been done using computer facilities of the Supercomputer Center, Institute of 
Solid State Physics, University of Tokyo.

\pagebreak

\end{document}